\documentclass{article}

\usepackage[utf8]{inputenc} 
\usepackage[T1]{fontenc}    
\usepackage{hyperref}       
\usepackage{url}            
\usepackage{booktabs}       
\usepackage{amsfonts, amsmath, amssymb}       
\usepackage{nicefrac}       
\usepackage{microtype}      
\usepackage{graphicx}       

\usepackage[left=1.2in, right=1.2in, top=1in, bottom=1in]{geometry}
\usepackage[style=chem-acs]{biblatex} 
\usepackage[labelfont=bf,textfont=md]{caption}
\usepackage{algorithm, algpseudocode}
\usepackage{authblk}
\usepackage{chemformula}

\title{Score dynamics: scaling molecular dynamics with picoseconds timestep via conditional diffusion model}
\author{Tim Hsu\thanks{hsu16@llnl.gov}}
\author{Babak Sadigh}
\author{Vasily Bulatov}
\author{Fei Zhou\thanks{zhou6@llnl.gov}}
\affil{Lawrence Livermore National Laboratory, Livermore, CA 94551, United States}
\date{}

\begin{document}
\maketitle
\begin{abstract}
We propose score dynamics (SD), a general framework for learning accelerated evolution operators with large timesteps from molecular-dynamics simulations. SD is centered around scores, or derivatives of the transition log-probability with respect to the dynamical degrees of freedom. The latter play the same role as force fields in MD but are used in denoising diffusion probability models to generate discrete transitions of the dynamical variables in an SD timestep, which can be orders of magnitude larger than a typical MD timestep. In this work, we construct graph neural network based score dynamics models of realistic molecular systems that are evolved with 10~ps timesteps. We demonstrate the efficacy of score dynamics with case studies of alanine dipeptide and short alkanes in aqueous solution. Both equilibrium predictions derived from the stationary distributions of the conditional probability and kinetic predictions for the transition rates and transition paths are in good agreement with MD. Our current SD implementation is about two orders of magnitude faster than the MD counterpart for the systems studied in this work. Open challenges and possible future remedies to improve score dynamics are also discussed.
\end{abstract}

\section{Introduction}
Molecular dynamics (MD) simulations are ubiquitous in condensed-matter physics, chemistry, materials science and biology. MD can be used to study equilibrium thermodynamic as well as kinetic properties of different phases of matter \autocite{Frenkel2002Understanding, Tuckerman2010Statistical}, thanks to its fine level treatment of all atomic details from first-principles, with no presumption other than the interatomic potential. However, such accuracy comes at high computational cost that severely limits the spatial and temporal scales accessible. Expanding these scales is key to unlocking MD's full predictive potential. 

Regarding the spatial scales, parallel computing has enabled simulations of molecular systems that can cross spatial scales and contain as many as $10^{10}$ atoms, which is gigantic compared to fewer than 100 at MD's inception \autocite{Alder1957}. Efforts to reduce the extravagant computational cost of MD at such large spatial scales have led to the development of strategies to reduce the number of degrees of freedom, often by disregarding irrelevant or rapidly oscillating dynamical variables and focusing instead on pertinent, slowly varying ones that sufficiently elucidate structural transformations as well as chemical reactions. A common example is implicit solvation techniques that exclude solvent molecules (e.g., water) altogether \autocite{Feig2004COSB}. Another non-exclusive approach is to design a small set of representative particles such as united atoms or beads in dissipative particle dynamics \autocite{Hoogerbrugge1992} and coarse-grained force fields \autocite{Chen2006JCP-comparison}.

Expanding the temporal scales is particularly important when the scientific question hinges on sampling rare events, which are ubiquitous. The temporal scales are even more challenging to expand: the intrinsically small timestep $\Delta t \approx 1$ fs is necessitated by the $\sim$THz frequency of typical phonon modes.  Since MD is based on time integration of the Newtonian equation of motion sequentially over such tiny $\Delta t$, scaling up the temporal scales of MD is not amenable to brute-force stacking of parallel hardware. The femtosecond timestep limit therefore places a tough constraint on the practical MD duration accessible within realistic wall-clock time, notwithstanding heroic demonstrations lasting for weeks or more. To deal with these limitations and facilitate rare event sampling, a large variety of methods have been devised \autocite{Perez2009}. Here we only mention a few without attempting a comprehensive review. On parallel hardware, parallel-replica dynamics can enhance rare event sampling with the assumption of transition state theory \autocite{Voter1998PRB-Parallel}. Methods such as dissipative particle dynamics \autocite{Hoogerbrugge1992} and coarse-grained molecular dynamics \autocite{Clementi2008COSB, Monticelli2008JCTC} take advantage of the correlation between the spatial and temporal scales and coarsen the particles being simulated by replacing fast degrees of freedom (e.g., light/fast atoms) with heavier/slower particles (e.g., polymer beads), resulting in lowered effective vibrational frequency compatible with a larger timestep. Some popular and productive ideas modify the potential energy surface (PES) or effective temperature to make barrier crossing more likely. Examples include hyperdynamics \autocite{Voter1997PRL-Hyperdynamics}, temperature-accelerated dynamics \autocite{Sorensen2000JCP-TAD}, Gaussian accelerated molecular dynamics, and metadynamics \autocite{Laio2002PNAS-metadynamics}. 

In contrast to MD-like methods based on differential equations, another class of methods mixes MD with Monte-Carlo (MC) methods to achieve more efficient configuration sampling, including replica-exchange MD \autocite{Sugita1999CPL-REMD}, force bias Monte Carlo \autocite{Neyts2013TCA} and hybrid MD/MC \autocite{Sadigh2012PRB-Scalable,Sadigh2012PRB-Calculation}. By interspersing MD steps that continuously evolve the atomic configuration and MC steps that introduce abrupt jumps to different energy basins, such hybrid methods  quickly sample rare events while keeping particle momenta in the simulated phase space. In further departure from MD, other methods such as kinetic Monte Carlo \autocite{Voter1985-KMC} and first-passage Monte Carlo \autocite{Opplestrup2006PRL-First-Passage} are formulated to promote ergodicity and efficient sampling of rare events. In these methods, momenta are ignored and one is concerned with atomic structures only.

\begin{figure}
    \centering
    \includegraphics[width=0.8\textwidth]{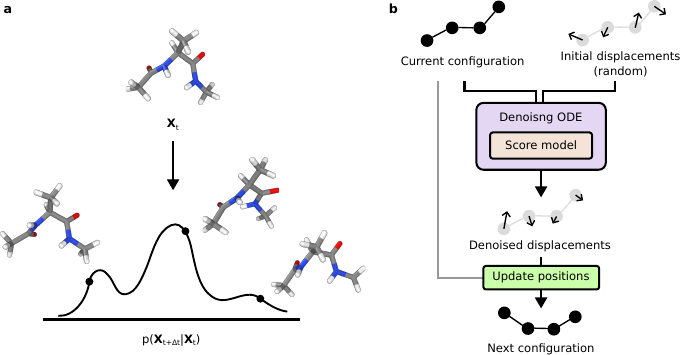}
    \caption{
    (a) Schematic illustration of distribution of diverging simulation outcomes after finite time interval.
    (b) Workflow of score dynamics, which maps both the current conditioning structure and a randomly displaced candidate structure into a new one by probabilistic denoising. 
    }
    \label{fig:intro}
\end{figure}

This work proposes \textit{score dynamics} (SD), an alternative machine-learning (ML) approach to dynamical simulation of atomistic systems. Recent years have witnessed the rise of scientific ML methods for atomistic simulations. Compared to the conventional method development paradigm of physical intuition and time-consuming trial-and-error, data-driven ML approaches have the potential to deliver quantitative models within relatively short development cycles based on large training data \autocite{Butler2018N, zhang2023artificial}. Score dynamics takes advantage of recent progress in ML generative models to sample discrete Markovian steps based on the transition probability matrix of molecular configurations over large timesteps (Fig.~\ref{fig:intro}a). At each step, conditioned on a current configuration, the next configuration is sampled via a conditional generative model, namely the score-based generative model a.k.a. denoising diffusion probabilistic model, or simply the diffusion model \autocite{Sohl-Dickstein2015-DPM, Ho2020-DDPM, Song2021-Score}, using a learned score function (Fig.~\ref{fig:intro}b). Both position-only and position+velocity versions of SD haven been developed, though we focus primarily on the former. Applied to alanine dipeptide and small alkanes as case studies, SD is shown to faithfully reproduce both the dynamical information of transition rates and transition paths, and the equilibrium distributions. Importantly, trained from a set of small alkanes not containing butane, the learned model is applicable to the unseen butane with accurate dynamical and equilibrium fidelity, demonstrating good generalizability akin to that of machine learning potentials.

\section{Methods} \label{sec:method}

\subsection{General formalism} \label{sec:formalism}

Before delving into the details of score dynamics, we start by motivating our design principles. The first one is a stochastic approach to dynamical simulations (Fig.~\ref{fig:intro}a) for the following three reasons. For the purpose of adopting large ($\gtrsim$ ps) timesteps $\Delta t$, the time-evolution of deterministic dynamical systems often diverge. Even infinitesimally close initial states $\mathbf{X}_0$ can and in most practical cases will evolve in divergent trajectories to a diverse set of states $\{\mathbf{X}_{\Delta t}\}$ of finite variance, as measured by the Lyapunov exponent. Here $\mathbf{X} =(\mathbf{r}, \mathbf{p})$ is the position-momentum state. Such a distribution of $\{\mathbf{X}_{\Delta t}\}$ is amenable to probabilistic treatment. For the purpose of spatial (particle) coarsening, the equation of motion of a smaller number of slow/relevant dynamical variables $\mathbf{A}$ projected from all atoms is ostensibly stochastic, as formulated under the Mori-Zwanzig formalism \autocite{mori1965transport, Zwanzig1973}:
\begin{align} \label{eq:Mori-Zwanzig}
    \frac{\partial \mathbf{A}(t)}{\partial t} = 
    i \Omega \mathbf{A}(t) - \sum_{0}^{t} K(s) \mathbf{A}(t-s) ds + F^R(t),
\end{align}
where $i \Omega$ is the frequency matrix, the memory kernel $K$ represents history-dependent effects \autocite{Klippenstein2021JPCB}, and $F^R$ is the contribution of fast/irrelevant variables being left out of explicit consideration. Even though the underlying all-atom Newtonian equation of motion is fully deterministic (assuming non-Langevin thermostat), the selective projection process will make the dynamics of the coarse variables appear to be affected by ostensibly ``random'' fluctuations, e.g. a solute molecules under the influence of ignored solvent molecules. Eq.~(\ref{eq:Mori-Zwanzig}) for the motion of $\mathbf{A}$ can be interpreted as a generalized Langevin equation. Note, however, practical closure of the Mori-Zwanzig formalism is difficult for complex systems \autocite{Li2015JCP-Incorporation, Li2014SM, Wang2020JCP, Klippenstein2021JPCB}. Finally, practical MD runs often adopt Langevin dynamics for thermostatting, making such simulations intrinsically stochastic. 

The second design decision is to adopt a discrete time approach, predicting the time series $\{\mathbf{X}_{n \Delta t}\}$ $(n=0,1,\dots)$ rather than relying on continuous-time ordinary differential equation (ODE, deterministic molecular dynamics) or stochastic differential equations (SDE, stochastic Langevin dynamics). Small timesteps are required for numerically stable  integration of differential equations. With large, discrete timesteps $\gg$ fs, score dynamics can achieve significant temporal scaling.

Given the above motivations, score dynamics is, in the most general setting, a high-order Markov-chain Monte Carlo (MCMC) method with a history-dependent conditional transition probability matrix $P(\mathbf{X}_{t'} | \mathbf{X}_{(0:n) \Delta t}, \Delta t, T)$
that describes the distribution of stochastic outcomes at the next step $t'=(n+1) \Delta t$ conditioned on the previous states $\mathbf{X}_{(0:n) \Delta t}$ at $t=0, \dots, n \Delta t$, satisfying
    $P(\mathbf{X} | \mathbf{X}_{(0:n) \Delta t}, \Delta t, T) \ge 0, \ \int P(\mathbf{X} | \mathbf{X}_{(0:n) \Delta t}, \Delta t, T) \text{d}\mathbf{X} = 1.$
SD makes autoregressive predictions by iteratively (1) drawing a random state $\mathbf{X}_{(n+1) \Delta t}$  according to $P(\mathbf{X}_{t'} | \mathbf{X}_{(0:n) \Delta t}, \Delta t, T)$, and (2) updating the conditioning states to $\mathbf{X}_{(0:n+1) \Delta t}$.

Generically formulated, SD is related to some existing approaches.
For example, with all atoms and infinitesimal $\Delta t$, SD is a single-step MCMC equivalent to Langevin dynamics, which is memoryless. More specifically, it is reduced in the $\Delta t \rightarrow 0$ limit to the Euler-Maruyama time integration of Langevin dynamics to $t'=t+\Delta t$
\begin{align}
    \text{d} \mathbf{p} &= 
        -( \nabla U + \gamma \mathbf{p} )\text{d} t + \sqrt{2  \gamma M k_B T } \text{d} \mathbf{w}_{t} \nonumber \\
    P(\mathbf{X}_{t'} | \mathbf{X}_{t}, \Delta t) &= 
        \mathcal{N}\left[ (\bar{\mathbf{r}}_{t'}, \bar{\mathbf{p}}_{t'}), \text{diag}(0, 2 \gamma M k_B T \Delta t) \right], \label{eq:Langevin} \\
    \bar{\mathbf{r}}_{t'} &= 
        \mathbf{r}_{t} +  \mathbf{p}_t \Delta t/M, ~ \bar{\mathbf{p}}_{t'} = \mathbf{p}_{t} - (  \nabla U + \gamma  \mathbf{p}_t )\Delta t, \nonumber
\end{align}
where $\gamma$, $U(\mathbf{r})$, $T$, $M$ are the damping coefficient, interatomic potential, temperature and mass, respectively, $\mathbf{w}_{t}$ is the standard Wiener process, and $\bar{x}_{t'}$ is the mean value at $t'$. The conditional $P(\mathbf{X}_{t'} | \mathbf{X}_{t}, \Delta t \rightarrow 0)$ is a Gaussian distribution with its mean and covariance matrix determined from the Langevin equation. Eq.~\ref{eq:Langevin} can be generalized from infinitesimal to any finite $\Delta t$: $P(\mathbf{X}_{t'} | \mathbf{X}_{t}, \Delta t)$ is formally solved by the Fokker-Planck Equation (FPE) in position-momentum space, also known as the Klein-Kramers equation (KKE), a PDE for evolution of the transition probability matrix $P(\mathbf{X}_{t'} | \mathbf{X}_{t}, \Delta t)$ over time interval $\Delta t$ \autocite{Risken1996-FPE}, though solving high-dimensional KKE is difficult in practice. Similar observations are made in the important limit of over-damped Langevin dynamics, or Brownian dynamics. In this regime, only the positions $\mathbf{r}=\mathbf{X}$ are kept, the Langevin equation becomes a first-order SDE for $\mathbf{r}$, and the corresponding transition probability can be solved from the position-only FPE, also known as the Smoluchowski equation \autocite{Risken1996-FPE}. In general, however, the dynamical variables are projected from the all-atom system, memory effects are relevant according to the Mori-Zwanzig formalism, and close-form solutions of transition $P$ are usually not available. In the large time limit $\Delta t \rightarrow \infty$, the conditional probability asymptotically becomes the stationary or equilibrium Boltzmann distribution $P_{\text{eq}}(\mathbf{r}) = \exp({-U(\mathbf{r})/k_B T})/Z$ independent of the conditioning states.

\textbf{Comparison with MD.} However, the goal to formally encapsulate all the complexities of dynamics over large spatiotemporal scales is fraught with technical difficulties. The transition probability $P(\mathbf{X}_{t'} | \mathbf{X}_{(0:n) \Delta t}, \Delta t, T)$ is significantly more challenging than and in many ways different from the interatomic potential $U(\mathbf{r})$ in MD. 
(1) The first major difference is that the initial conditioning states $\mathbf{X}_{(0:n) \Delta t}$ and final $\mathbf{X}_{t'}$ play equally important roles in the bivariate $P(\mathbf{X}_{t'} | \mathbf{X}_{(0:n) \Delta t}, \Delta t, T)$ (Fig.~\ref{fig:intro}b), in contrast to the univariate potential energy  $U(\mathbf{r})$ taking a single input structure.
(2) The former needs to be adjusted or retrained when a different timestep or temperature is applied, while the latter is universal and can be trained once and reused. 
(3) We will train the SD model with self-supervised learning of the complex dynamics \autocite{Yang2021P-Self} using high-fidelity MD trajectories as training data, i.e.\ forcing the model to predict the next configuration from current configurations (see next section for details), instead of supervised learning with pre-calculated total energies and forces as training labels for fitting potential $U$.
(4) SD's training data, MD trajectories, are intrinsically noisy with significant thermal fluctuations, especially at high temperature, whereas MD can be trained with clean, well-converged total energy and forces from e.g.\ quantum-mechanical calculations. 
(5) SD as a probabilistic model needs to be able to provide statistical information including mean and variance of the predicted next configuration. For example, to quantify the statistical distribution of configurations $\mathbf{X}_{t'}$ starting from the same initial $\mathbf{X}_{0}$, \textit{many} independent runs need to be repeated to establish statistical estimates. Therefore, SD  requires significant amount of MD trajectory data to be accurate, the more the better. In contrast,  potentials are deterministic, requiring a single set of labels (energy and forces) for each input configuration.  For all these reasons, a major drawback of score dynamics is high up-front training cost. We expect the number of MD snapshots for SD training to be several orders of magnitude higher than MD.
(6) Finally, even if the transition probability $P(\mathbf{X}_{t'} | \mathbf{X}_{(0:n) \Delta t}; \Delta t; T)$ is exactly known, sampling $\mathbf{X}_{t'}$ from this high-dimensional distribution  is far from trivial.

There is relief. Note that MD directly requires forces, or derivative of potential energy $\mathbf{F}=-\partial U/\partial \mathbf{X}$. While not obvious, a similar statement can be made for SD: the \textit{score} function, here defined as the derivative of conditional log-probability 
\begin{equation} \label{eq:score}
    \mathbf{s}(\mathbf{X}_{t'}, \mathbf{X}_{(0:n) \Delta t}, \Delta t, T) =
        \frac{\partial \log P(\mathbf{X}_{t'} | \mathbf{X}_{(0:n) \Delta t}; \Delta t; T)} {\partial \mathbf{X}_{t'}},
\end{equation}
can be used to sample new configurations $\mathbf{X}_{t'}$ \autocite{Sohl-Dickstein2015-DPM, Ho2020-DDPM, Song2021-Score} (details in the following section). This is a significant simplification: a properly defined probability $P(\mathbf{X}_{t'} | \mathbf{X}_{(0:n) \Delta t}; \Delta t; T)$ requires normalization, a notoriously complex and expensive process involving integration over the whole high-dimensional phase space of configurations $\mathbf{X}$. Scores circumvent the need for intractable normalization. The accuracy of SD in practice therefore hinges on the quality of the approximate score model $\mathbf{s}_\theta$. The rest of the paper will discuss how to approximate scores $\mathbf{s}$ with ML models $\mathbf{s}_\theta$.

\subsection{Approximations by physical considerations} \label{sec:approx}

So far we presented score dynamics as an exact, finite-timestep reformulation of MD with no loss of accuracy. It is still not useful unless we can accurately approximate the scores by learning from small-scale MD and extrapolate to larger systems, just like the fact that MD potentials and force fields are typically trained from a small number of atoms. To do this, likewise, simplifications have beeen introduced to make score-fitting tractable. 
The first approximation in this work is to ignore history dependence and adopt a one-step MCMC approach, which is considerably simpler than multi-step. Second, we assume the overdamped regime and ignore velocities. With a large timestep $\Delta t=10$ ps, the velocity of solute molecules in our case studies are essentially uncorrelated \autocite{Daldrop2018PNAS}, leaving $\mathbf{X} =\mathbf{r}$. This can be shown quantitatively in the velocity autocorrelation function (VAF) of alanine dipeptide in Supplemental Fig.~4a, which shows that the velocities at 1 ps intervals are essentially uncorrelated. To test the validity of the position-only approximation, we trained a separate score dynamics model in the full position+velocity phase space, which reproduces vanishing VAF (red dot in Supplemental Fig.~4a). We focus on position-only SD in this work thereupon. Third, our score model directly predicts the score rather than taking the derivative of $\log P$ in Eq.~(\ref{eq:score}), similar to fitting a force-field without a scalar potential. We also limited ourselves to fixed $\Delta t=10$ ps and $T=300$ K. With these approximations and simplifications, the score is $\mathbf{s}(\mathbf{r}_{t'}, \mathbf{r}_{t}, \Delta t, T)$. The validity of these approximations is ultimately determined by the resulting accuracy. Empirical tests show that the simplified score dynamics with position-only MCMC is accurate enough for the relatively simple molecules considered in this work.

While all high-fidelity training MD simulations include all atoms, spatial (particle) coarsening will be demonstrated with an implicit-solvation-like model with solute molecule only and  water molecules are omitted from explicit consideration.

\subsection{Conditional diffusion model}

In this section, we rewrite the transition probability matrix $P(\mathbf{r}_{t'}| \mathbf{r}_{t}, \Delta t, T) =q(\mathbf{z}| c)$ as a conditional distribution $q$ of displacement $\mathbf{z} = \mathbf{r}_{t'} - \mathbf{r}_t$ for brevity. Here $c=(\mathbf{r}_{t}, \Delta t, T)$ is a short-hand for the conditioning variables.
The key technical approach of score dynamics is to approximate the corresponding score $\mathbf{s}(\mathbf{z}| c)$ using the diffusion model 
\autocite{Sohl-Dickstein2015-DPM, Ho2020-DDPM, Song2021-Score}. 
The diffusion model establishes a mapping between the real data samples $\mathbf{z}$ and a latent distribution $p'(\mathbf{z}')$---which is typically independent and identically distributed (iid) Gaussian---by iteratively applying Gaussian noises to $\mathbf{z}$ towards $p'(\mathbf{z}')$. The generation of realistic $\mathbf{z}$ is then achieved by sampling a latent  $\mathbf{z}'$ and mapping it back to the original data space. 
The main advantage of the diffusion model over other generative models is the overall superior performance in sampling high-dimensional data such as images and videos, even though at the cost of higher computational burdens \autocite{yang2023diffusion}.
The diffusion model derives its name from a forward and a backward diffusion process, or auxiliary Langevin dynamics in fictitious time $\tau$ (not to be confused with the physical time $t$).
The \textit{forward noising} process maps data to code:
\begin{align}
    \text{d} \mathbf{z}_{\tau} &= f(\tau)\mathbf{z}_{\tau} \text{d} \tau + g(\tau) \text{d} \mathbf{w}_{\tau}, \nonumber \\ 
    \ q_{\tau}(\mathbf{z}_{\tau}|\mathbf{z}_{0}) &= \mathcal{N}(\mathbf{z}_{\tau}|\alpha_{\tau} \mathbf{z}_{0}, \sigma^2_{\tau} \mathbf{I} ), 
    \label{eq:Forward-noising} \\
    f(\tau) &= \frac{\text{d} \log \alpha_{\tau}} {\text{d} \tau}, ~ 
    g^2(\tau) = \frac{\text{d} \sigma^2_{\tau}} {\text{d} \tau} - 2 \sigma^2_{\tau} \frac{\text{d} \log \alpha_{\tau}} {\text{d} \tau}, \nonumber
\end{align}
where $(\alpha_\tau, \sigma_\tau)$ is the so-called noise schedule, $f(\tau)$ and $g(\tau)$ are related drift and diffusion coefficients, and $\mathbf{w}_{\tau}$ is the standard Wiener process.The span of $\tau$ is typically set to $(\tau_{\min}\sim 0^+, \tau_{\max}=1)$, over which $\alpha_\tau$ monotonically decreases from 1 to 0, while $\sigma_\tau$ monotonically increases from 0 to 1. Following this noise schedule, it can be seen that $\mathbf{z}_\tau$ initially is the ``clean'', original data $\mathbf{z}=\mathbf{z}_0$ at $\tau \rightarrow 0$, and approaches iid Gaussian at $\tau \rightarrow 1$.

While the forward noising process is trivial, the \textit{reverse denoising} operation, or the code-to-data mapping, is nontrivial and learned by a machine learning model. It can be formulated as an SDE (not shown here) or, more convenient for our purpose, a probability-flow ODE \autocite{Song2021-Score}:
\begin{align}
    \frac{\mathrm{d} \mathbf{z}_\tau} {\mathrm{d} \tau} 
        & =       f(\tau) \mathbf{z}_\tau -\frac{1}{2} g^2(\tau) \nabla_{ \mathbf{z}_\tau}  \log q_\tau(\mathbf{z}_\tau|c) \nonumber \\
        & \approx f(\tau) \mathbf{z}_\tau -\frac{1}{2}            g^2(\tau) s_{\theta }       (\mathbf{z}_\tau ,\tau, c)   \nonumber \\
        & =       f(\tau) \mathbf{z}_\tau +\frac{1}{2\sigma_\tau} g^2(\tau) \epsilon_{\theta }(\mathbf{z}_\tau ,\tau, c)  \label{eq:deoise-ODE},
\end{align}
where the score function $\nabla_{ \mathbf{z}_\tau} \log q_\tau(\mathbf{z}_\tau|c)$ is approximated by a score model $s_{\theta}$ with parameters $\theta$, or a noise-prediction network $\epsilon_{\theta}=- s_{\theta} \sigma_\tau$ to predict the noise $\epsilon$ in the noisy $\mathbf{z}_\tau$ from Eq.~(\ref{eq:Forward-noising}). 
The conditioning variable $c$ is kept throughout to emphasize the dependence on input $c$.
Starting with $\mathbf{z}_{\tau}$ sampled from iid Gaussian at $\tau = 1$, Eq.~(\ref{eq:deoise-ODE}) can be integrated back to $\tau=0$ to generate the desired displacement $\mathbf{z}=\mathbf{z}_0$. 
Operationally, score dynamics replaces MD's physical time integration of $\Delta t/1 \text{fs} \approx 10^4$ steps with a fictitious time integration of $N_{\tau}$ steps between $\tau_{\max}$ and 0. Small $N_{\tau}$ is needed for SD to be competitive and worthwhile.
The recent DPM-Solver \autocite{lu2022dpm} recognizes that the form of the denoising ODE can be solved by a more dedicated exponential integrator algorithm, thereby achieving realistic sample generation in a limited number of denoising timesteps $(N_{\tau}< 100)$. We used the \textit{solver3} variant of DPM-Solver to solve Equation \ref{eq:deoise-ODE} for generating displacement $\mathbf{z}$, in typically $N_{\tau}\approx$ 8--20 auxiliary timesteps, or 24--60 function evaluations of $\epsilon_\theta$. Inference or time evolution prediction using score dynamics is summarized in Algorithm \ref{alg:infer}.

\begin{algorithm}
\caption{Score dynamics simulation} \label{alg:infer}
\begin{algorithmic}    
    \Require{Initial $t=0$, configuration $\mathbf{r}_0$,
        time step $\Delta t$, temperature $T$,
        score model $\epsilon_\theta$,
        and noise schedule $(\alpha_\tau, \sigma_\tau)$
        }
    \Repeat
        \State $c \gets (\mathbf{r}_t, \Delta t, T)$
        \State $\tau \gets \tau_{\max}$
        \State Sample $\mathbf{z}_{\tau} \sim \mathcal{N}(0, \sigma_{\tau}^2 \mathbf{I})$
        \State $\mathbf{z}_{0} \gets 
        \mathbf{z}_{\tau_{\max}} - \int_{\tau_{\min}}^{\tau_{\max}} \text{d} \tau \left[f(\tau) \mathbf{z}_\tau +\frac{1}{2\sigma_\tau} g^2(\tau) \epsilon_{\theta }(\mathbf{z}_\tau ,\tau, c) \right] $
        \State $\mathbf{r}_{t+\Delta t} \gets \mathbf{r}_t + \mathbf{z}_0$
        \State $t \gets t+\Delta t$
    \Until{end of simulation}
\end{algorithmic}
\end{algorithm}

\subsection{Score model training} \label{sec:training}

The noise-prediction network $\epsilon_{\theta}(\mathbf{z}_{\tau}, \tau, \mathbf{r}_t, \Delta t,  T)$ is trained from MD trajectory data at the desired interval $\Delta t$. In this work, we fixed the timestep $\Delta t$ and temperature $T$, so they are implicit. The network is then written as $\epsilon_{\theta}(\mathbf{z}_\tau, \tau, \mathbf{r}_t)$ with three inputs: noisy displacement $\mathbf{z}_{\tau}$, fictitious time $\tau$, and conditioning structure $\mathbf{r}_t$. We convert $\mathbf{r}_t$ and $\mathbf{z}_{\tau}$ to a combined molecular graph, which is then subject to graph neural network (GNN) operation. Given a MD trajectory $\{\mathbf{r}_{t} \}$, each data point in the training set $\mathcal{D}$ is $(\mathbf{r}_{ t}, \mathbf{z}= \mathbf{r}_{t+\Delta t}-\mathbf{r}_{t})$. The network is trained with the denoising score-matching loss \autocite{Vincent2011scorematching}
\begin{align} \label{eq:loss-DSM}
    \mathcal{L}_{\text{DSM}} = 
    \mathbb{E}_{\mathbf{r}_{t}, \mathbf{z}_0, \epsilon, \tau} 
        \Vert \epsilon - \epsilon_{\theta}(\mathbf{z}_\tau, \tau, \mathbf{r}_t) \Vert^2.
\end{align}

Combining the description of the training data, the forward noising process, and the loss function, Algorithm \ref{alg:train} lists the pseudocode for training the noise model $\epsilon_{\theta}$. Exponential moving average (EMA) was used to facilitate model training. Rectified Adam (RAdam) was used as the gradient descent optimizer without learning rate decay or weight decay. The noise schedule $(\alpha_\tau, \sigma_{\tau})$ follows the variance-preserving linear variant  \autocite{Song2021-Score}: 
\begin{align*}
    \log \gamma_\tau = -\frac{\beta_1-\beta_0}{2} \tau^2 - \beta_0\tau, \ \alpha_\tau= b \sqrt{\gamma_\tau}, \ \sigma_\tau=\sqrt{1- \gamma_\tau},
\end{align*}
where $b=2$ is a simple scaling factor for appropriate level of signal-to-noise ratio, $\beta_0 = 0.1$ and $\beta_1 = 20$. For every sampled minibatch of the molecules, the atomic coordinates $\mathbf{r}$ are randomly rotated as part of the data augmentation process.

\begin{algorithm}
\caption{Score model training} \label{alg:train}
\begin{algorithmic}    
    \Require{Training dataset $\mathcal{D}$,
        noise schedule $(\alpha_\tau, \sigma_\tau)$,
        initial model parameters $\theta$,
        model parameters $\theta'$ for EMA,
        EMA decay rate $\beta$,
        data augmentation $\mathcal{A}$,
        gradient descent optimizer Optim,
        and learning rate $\eta$.
    }
    \Repeat
        \State Sample $(\mathbf{r}_t, \mathbf{z}_0=\mathbf{r}_{t+\Delta t}-\mathbf{r}_t) \sim \mathcal{D}$
        \State Sample $\tau \sim \mathcal{U}(\tau_{\min}, \tau_{\max})$
        \State Sample $\epsilon \sim \mathcal{N}(0, \mathbf{I})$
        \State $\mathbf{r}_t, \mathbf{z}_0 \gets \mathcal{A}(\mathbf{r}_t, \mathbf{z}_0)$
        \State $\mathbf{z}_\tau \gets \alpha_\tau \mathbf{z}_0 + \sigma_\tau \epsilon$
        \State $\mathcal{L} \gets \mathbb{E}_{\mathbf{r}_{t}, \mathbf{z}_0, \epsilon, \tau} 
        \Vert \epsilon - \epsilon_{\theta}(\mathbf{z}_\tau, \tau, \mathbf{r}_t) \Vert^2
    $
        \State $\theta \gets \mathrm{Optim}(\mathcal{L}, \theta, \eta)$
        \State $\theta' \gets \beta \theta' + (1-\beta)\theta $
    \Until{convergence}
\end{algorithmic}
\end{algorithm}

The training parameters are detailed in Table \ref{tab:train-params}. All other parameters, if unspecified, default to implementations in PyTorch 1.11.0 \autocite{paszke2019pytorch} and PyTorch-Geometric 2.0.4 \autocite{fey2019fast}, which were used for the GNN implementation and its training.

\subsection{Score model architecture}
The noise-prediction GNN consists of three components: an Encoder, a Processor for graph message-passing, and a Decoder. The input molecular graph to these operations is denoted as $(V, E)$, where $V$ consists of node features $\mathbf{h}_i$ for the $i$th node (or atom), and $E$ consists of edge (or bond) features $\mathbf{e}_{ij}$ from node $i$ to node $j$. In contrast to conventional molecular graph representations applied to interatomic potential or molecular property predictions with a single input structure $\mathbf{r}=\mathbf{r}_t$, our representation needs to incorporate two extra variables: the proposed structure $\mathbf{r'}=\mathbf{r}_{t+\Delta t}$ and fictitious time $\tau$. This is implemented using a special graph Encoder. The node feature $\mathbf{h}_i$ combines a one-hot embedding of the atom/node type $a$ followed by a simple multi-layer perception (MLP) $M_H$  and an embedding of fictitious time $\tau$ by Gaussian Fourier features (Gaussian RFF) \autocite{tancik2020fourier} and an MLP $M_T$:
\begin{equation} \label{eq:encoding} 
    \mathbf{h}_i \gets M_H(\text{onehot}(a_i)) + M_T(\text{GaussianRFF}(\tau)).
\end{equation} 
The set of edges $E$ is determined by a cutoff distance $r_c$ according to the conditioning structure $\mathbf{r}$, within which a pair of nodes form an edge connection. The edge feature $\mathbf{e}_{ij}$ is the concatenation of the conditioning bond vectors $\mathbf{r}_{ij} = \mathbf{r}_{j} - \mathbf{r}_{i}$ at $t$ and the bond vectors after noisy displacements $\mathbf{r}'_{ij} = \mathbf{r}_{ij}  + \mathbf{z} _{\tau, j} - \mathbf{z}_{\tau,i}$ at fictitious time $\tau$, followed by an MLP $M_E$:
\begin{equation}
    \mathbf{e}_{ij} \gets M_E( 
            \mathbf{r}_{ij} \oplus {r}_{ij} \oplus \mathbf{r}'_{ ij} \oplus {r}'_{ ij}
        ),
\end{equation}
where ${r}_{ij}$ and ${r}'_{ij}$ represent distance, $\oplus$ designates feature concatentation, $M_H$, $M_E$ and $M_T$ consist of two dense linear layers, with SiLU activation function \autocite{Hendrycks2016-GELU} after the first layer and layer normalization after the second.

We used the Processor from MeshGraphNets \autocite{Pfaff2020} for implementing graph convolutions. Although we have considered rotationally equivariant GNNs for predicting the noise $\epsilon$ (a vector quantity), the simple design of MeshGraphNets renders the computation to be fast without much loss to accuracy, as demonstrated in previous surrogate models for dislocation \autocite{bertin2023accelerating, bertin2023learning} and microstructure evolution \autocite{Fan2023-Accelerate}. The computational speed is an important aspect when performing long rollouts that would take hundreds of thousands of function evaluations. Of course, the choice of a non-equivariant model means that random rotational data augmentation is necessary at training time. The Processor updates edge features $\mathbf{e}_{ij}$ and the node features $\mathbf{h}_i$ with MLPs $M_{PE}$ and $M_{PO}$ that have the same architecture as $M_H$, $M_E$, and $M_T$:
\begin{equation} \label{eq:message-passing}
    \mathbf{e}^{\prime}_{ij} \leftarrow 
        \mathbf{e}_{ij} + M_{PE}(
            \mathbf{e}_{ij} \oplus \mathbf{h}_i \oplus \mathbf{h}_j
        ), ~
    \mathbf{h}^{\prime}_{i}  \leftarrow 
        \mathbf{h}_i + M_{PH}(
            \mathbf{h}_i \oplus \sum_{j \in \mathcal{N}(i)} \mathbf{e}^{\prime}_{ij}
        ),
\end{equation}
where $\mathcal{N}(i)$ stands for the neighbor nodes of node $i$. Eq.~(\ref{eq:message-passing}) constitutes one graph convolution or message-passing layer, and the Processor consists of stacking or composition of such layers into a deep graph network. We refer to the original paper for further details of MeshGraphNets \autocite{Pfaff2020}. Note that in the original MeshGraphNets work, there are two sets of edges, whereas there is only one ($\mathbf{e}_{ij}$) in our work.

The Decoder is a simple MLP $M_O$ that maps the node features at the last layer into the predicted noise:
\begin{equation}
    \epsilon_i \gets M_O(\mathbf{h}_i).
\end{equation}
$M_O$ has the same architecture as that of $M_H$, $M_E$ and $M_T$, except that there is no layer normalization at the end. 

The model parameters are listed in Table \ref{tab:train-params}.

\begin{table}
    \centering
    \caption{Model and training hyperparameters.}
    \begin{tabular}{l l}
    \toprule
    Name & Value \\
    \midrule
    Number of graph convolutions        & 5\\
    Node, edge, time feature dimensions & 200\\
    Edge cutoff $r_c$                   & 4 \AA \\
    $(\tau_\mathrm{min}, \tau_\mathrm{max})$  & (0.001, 1) \\
    Number of model updates             & 800k\\
    Learning rate $\eta$                & 0.001\\
    EMA decay rate $\beta$              & 0.999      \\
    Minibatch size                      & 256        \\
    \bottomrule
    \end{tabular}
    \label{tab:train-params}
\end{table}

\subsection{Related works}
Machine-learned interatomic potentials, one of the best known ML applications to atomistic modeling, have attracted tremendous interests and investments, and are becoming powerful potential energy approximations that promise quantum mechanical accuracy at vastly reduced computational costs \autocite{Behler2016JCP-MLIAP}. Meanwhile, deep generative methods, including variational autoencoders, normalizing flow, and diffusion model, have also been applied for protein folding prediction \autocite{Wu2021.06.06.447297}, static structure generation \autocite{Noe2019S-Boltzmann-generator, Xu2022-Poisson-flow, Xie2022-dpm-crystal, Xu2022-dpm-molecule, Jo2022, zheng2023towards}, and collective variable analysis \autocite{Wang2022PNAS}. Employing the denoising score-matching algorithm on perturbed atomic structures, we recently developed a score-based denoising method for identification of crystal phases and defects with state-of-the-art accuracy \autocite{Hsu2022-denoiser}. In comparison, the development of ML methods to accelerate MD simulations is overall in a relative nascent stage that is quickly expanding and growing with promising new ideas and results. Similar to MD potential development, coarse-grained simulators can benefit from more accurate representations with ML \autocite{Wang2019ACS, Doi2023-ML-DPD}. In the past few years, probabilistic generative methods, including most notably normalizing flow \autocite{Liu2022SJNA, Klein2023-Timewarp} and  diffusion models \autocite{Fu2022, Wu2022, Arts2023, Lu2023-SamplingProtein, Schreiner2023-Implicit}, are being applied towards surrogate simulators trained with ground-truth data from high-fidelity trajectory. Liu and co-workers employed normalizing flows as the deep generative model to drawn samples of the Fokker-Planck equation \autocite{Liu2022SJNA}. Timewarp, a recent MCMC method achieved significantly accelerated thermodynamic sampling of the configurational space of small peptides using a conditional normalizing flow model for proposing new moves combined with parallel replicas and Metropolis algorithm. Good agreement with the equilibrium distribution and tranferability to unseen structures were shown for small peptide molecules, though at the cost of deviating from the ground truth kinetics. Timewarp was also demonstrated with even higher efficiency as a tool for configuration exploration in an alternative rejection-free mode, i.e.\ without enforcing equilibrium thermodynamics \autocite{Klein2023-Timewarp}. Fu \textit{et al.} applied a graph clustering algorithm to train coarse-grained force fields that can be used in molecular dynamics with very large time steps ($\sim$ns), and optionally adopted a conditional diffusion model to refine the coarse-grained molecular dynamics and bring the simulated structures closer to physical or likely ones. While their conditional diffusion models are similar to this work, the time-evolution simulation is carried out by molecular dynamics. Wu and Li proposed DiffMD, a diffusion model for trajectory prediction, with the noted distinction of injecting the current configuration into the noise level \autocite{Wu2022}. Arts {\textit{et al.}} developed a coarse-grained force field using a diffusion model for accelerated molecular (Langevin) dynamics \autocite{Arts2023}. Lu {\textit{et al.}} enhanced the efficiency of sampling protein conformations by large random displacements followed by structural refinement with a diffusion model  \autocite{Lu2023-SamplingProtein}. Schreiner \textit{et al.} proposed Implicit Transfer Operator, a multi-resolution surrogate model for protein folding implemented using a diffusion model with large, variable timesteps \autocite{Schreiner2023-Implicit}.

\subsection{MD simulations} \label{sec:MD}
MD Simulations are performed with the GROMACS (Version 2022.2) \autocite{Abraham2015S-Gromacs} package with mixed precision and CUDA support. Each case study consists of a single molecule (alanine dipeptide or small alkane molecule) solvated in water, and considers only intra-molecular effects. Atomic interactions are parametrized with the OPLS all-atom force field \autocite{Jorgensen1996JACS-OPLS} and solvated in a periodic cubic box with about 3000 TIP4 water molecules \autocite{Jorgensen1983JCP-TIP4P}.  The Linear Constraint Solver (LINCS) algorithm was used to enforce bond constraints \autocite{LINCS}. The Particle-mesh Ewald scheme \autocite{PME} was used to treat long-range electrostatics with a cutoff radius of 1.4 nm. After energy minimization of the initial solvated structure, the cubic cell size was equilibrated with NPT simulations at 1 bar for 0.2 ns using the the Berendsen barostat \autocite{Berendsen1984JCP-Molecular}, and all subsequent simulations were performed in the NVT ensemble. All runs were integrated with Langevin dynamics at 300 K with a timestep of 1 fs and a friction coupling time of 0.1 ps unless otherwise noted. The validation trajectory includes 120 ns. The MD speed is 247 wall-clock hours per $\mu$s of MD simulation using 1 IBM-Power9 CPU core and 1 NVidia V100 GPU, or 101 wall-clock hours on 14 cores and 4 GPUs.

{\textbf{Training set.}} All training MD trajectories were recorded at 1 ps intervals. During training, pairs of configurations were randomly sampled at recorded frame $t$ and $t+10$, i.e.\ interval of 10 ps.  An NVT trajectory of 100 ns was included for each molecule. To add a more diverse set of initial configurations in the training set, each alkane molecule was simulated in vacuum at a very high temperature of $10^4$ K at a timestep of 0.5 fs for $4\times 10^6$ steps with 4000 high-temperature alkane structures saved at 1000 step intervals. Each of the 4000 was solvated in water and equilibrated under NVT for 20 ps with free water and fixed solute molecules, followed by production NVT runs for 8 ps. The 4000 shorter trajectories were added to the long NVT training trajectory for each alkane molecule. Adding diverse initial structures for alanine dipeptide was done by randomly sampling the dihedral angles $\phi$ and $\psi \in [0, 2\pi]$ in each of 6000 initial structures, followed by similar solvation and equilibration procedures, as well as final short NVT training runs for 20 ps. The long and short NVT training MD runs have the same cell size and number of water molecules.

\section{Results} \label{sec:results}

We attempt to accurately approximate the scores by learning from small-scale MD, just like the fact that MD potentials and force fields are typically trained from a small number of atoms. In this work, we begin with relatively simple molecules in aqueous solutions, including alanine dipeptide and small alkanes  at fixed timestep $\Delta t=10$ ps and temperature $T=300$ K, with a focus on detailed analysis of the performance of score dynamics compared to the ground truth MD simulations. A generalization test (to \textit{unseen} butane) was also performed. Applications to larger molecular systems will be investigated in future work.

\subsection{Rejection criteria and resampling} \label{sec:reject-criteria}

The sampling procedure of score dynamics described in Algorithm \ref{alg:infer} is rejection-free if the trained score model is exact. However, in practice, unphysical configurations are found to be sampled in very rare cases, e.g.\ breaking of C-C covalent bonds, due to inexact scores. To maintain the stability of a very long simulation is always a challenge for dynamical methods. We mitigate this by simply rejecting such outlier samples based on some pre-defined criteria, followed by resampling. Accordingly, the raw transition matrix $P$ is replaced by
\begin{align} \label{eq:rejection}
    \tilde{P}(\mathbf{r}'| \mathbf{r}) &= P(\mathbf{r}'| \mathbf{r}) *A(\mathbf{r}'),
\end{align}
where the acceptance $A(\mathbf{r}')$ is 1 if the proposed new $\mathbf{r}'$ satisfies the predefined criteria and 0 otherwise. This make the SD implementation presented in this work a Monte-Carlo method with very rare rejections (typically less than 0.01\%).

In the current work the criteria involves checking the distances between the bonded pairs to ensure that no ``bond-breaking'' occurs. Additionally, for alanine dipeptide, which is studied in this work, the chirality is also monitored. Such evaluations have negligible computational cost relative to the diffusion model sampling process and is observed to not be a bottleneck in dynamic rollouts.
The details of the bond length criteria implemented in this work are as follows:
for alanine dipeptide, the range of (0.9, 1.16) \AA, in angstroms, for X-H bond lengths (where X is either of \{C, N, O\}), and (1.1, 1.7) for X-X bonds; and for alkanes, (1.02, 1.16) for C-H bonds and (1.38, 1.70) for C-C bonds. More quantitative results on the frequency of rejection will be given later.

\subsection{Alanine dipeptide}

\begin{figure}
    \centering
    \includegraphics[width=1.0\textwidth]{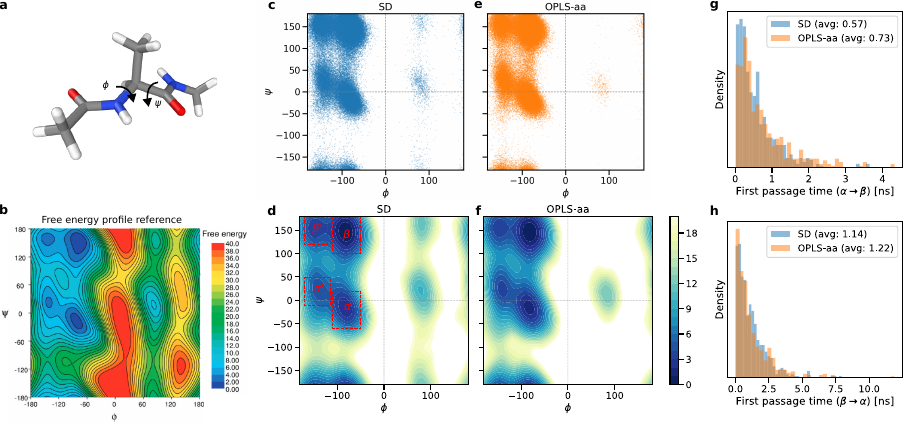}
    \caption{The alanine dipeptide SD trajectory is realistic relative to literature reference and in-house MD data.
    (a) The alanine dipeptide structure and the dihedral angles $(\phi, \psi)$ used for describing the instantaneous configurations.
    (b) Free energy profile reference, reprinted (adapted) with permission from \autocite{vymetal2010metadynamics} (copyright 2010 American Chemical Society).
    (c, d) SD trajectory (over 100ns with 1ps timesteps) visualized in Ramachandran plots, in both scatter and KDE (kernel density estimation) energy contour format. Two energy minima are annotated as $\alpha$ and $\beta$.
    (e, f) In-house MD trajectory reference based on the OPLS-aa potential visualized in Ramachandran plots.  
    (g, h) First passage time ($\alpha \rightarrow \beta$ and $\beta \rightarrow \alpha$) distribution from the SD trajectory compared to the in-house MD reference.
    The energy levels in (b, d, f) are in the unit of kJ/mol.
    }
    \label{fig:ala-dipep}
\end{figure}

Trained with MD trajectories of alanine dipeptide, a common molecule for prototyping, the corresponding SD trajectory is largely faithful to the MD training set (Fig.~\ref{fig:ala-dipep}). Specifically, the equilibrium dihedral angle (see Fig.~\ref{fig:ala-dipep}a) distribution or Ramachandran plot (c, d), measured over a 100 ns trajectory length, matches a literature reference PES \autocite{vymetal2010metadynamics} in the low-energy regions (b) and our ground-truth MD results (e, f). Note that the higher energy regions in (b) were computed from metadynamics \autocite{vymetal2010metadynamics} and too rare to be accessible from MD or SD simulations (c, e, d, f) within 100 ns. Further, the dynamical information, quantified as the histogram of the first-passage time between two energy minima (labeled $\alpha$ and $\beta$) extracted from the same 100 ns trajectory, are also in quantitative agreement with ground-truth MD (g, h). In the current work, the SD trajectory tends to be slightly more ``jumpy'', or more likely to overcome energy barriers, than the MD reference. This can be observed in (c, d) where high energy configurations around $\phi=90^{\circ}$ were oversampled compared to the reference MD in (e, f), and in (g, h) where the mean first passage time is slightly less than that of the MD reference. There is indeed a systematic slight underestimation of the energy barriers by the trained score model. The origin of this discrepancy in the high energy regions is currently unclear and will be investigated in future works. Possible speculative reasons include deficiency in the score model architecture or training dataset.

\begin{figure}
    \centering
    \includegraphics[width=0.8\textwidth]{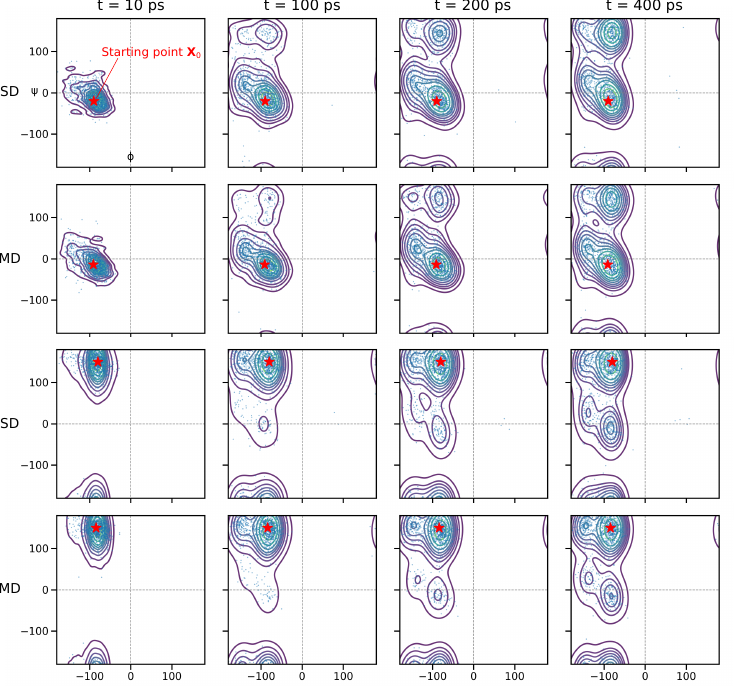}
    \caption{Ramachandran plots of conditional distributions $P(\mathbf{X}_t | \mathbf{X}_0)$ from SD trajectories and in-house MD data. In each plot, 1,000 scatter points (from 1,000 independent runs) and their KDE contours are plotted.
    }
    \label{fig:ala-dipep-cond-dist}
\end{figure}

We also compared the SD result with MD in terms of the conditional distribution $P(\mathbf{X}_t | \mathbf{X}_0)$, which is empirically measured by running 1,000 independent trajectories starting with the same initial configuration $\mathbf{X}_0$. This is straightforward in SD: simply rerun from the same initial condition with different random seeds. In MD, we first fixed the atoms on the dipeptide while allowing the water molecules to equilibrate, and then restarted with random equilibrium velocities on all atoms. As shown in Fig.~\ref{fig:ala-dipep-cond-dist}, the SD conditional distributions (described by the dihedral angles $\psi$ and $\phi$), either from the energy minimum $\alpha$ or from $\beta$ as the starting point, are similar to that of the MD reference data. Good agreement between SD and MD is observed with increasing simulation time $t$ = 10, 100, 200, and 400 ps, and both methods gradually approach the equilibrium distribution, as expected. Based on this result, SD is therefore reasonably accurate in terms of the conditional distributions $P(\mathbf{X}_t | \mathbf{X}_0)$ over a wide range of $t$ and up until $t$ is significantly large such that the conditional distribution approaches the equilibrium distribution $P(\mathbf{X}_t | \mathbf{X}_0) \rightarrow P_{\text{eq}}(\mathbf{X})$. 

\begin{figure}
    \centering
    \includegraphics[width=0.8\textwidth]{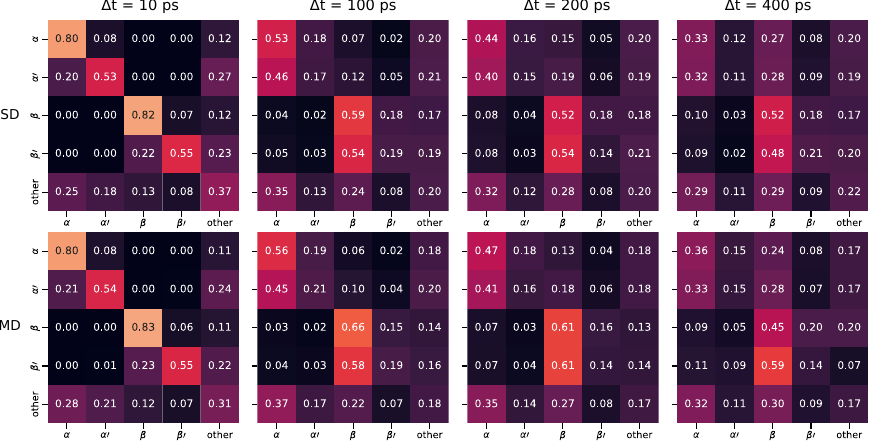}
    \caption{Transition probability matrix $P_{\Delta t}(C_j | C_i)$ for alanine dipeptide SD and MD trajectories, where $C$ are classes of dihedral distributions labeled as $\alpha$, $\alpha'$, $\beta$, $\beta'$, and `other', as shown in Fig.~\ref{fig:ala-dipep}d. Subscript $i$ corresponds to rows of the matrix, and $j$ corresponds to columns. This data was measured from the 1,000 independent trajectory runs (starting from $\alpha$) shown in Fig.~\ref{fig:ala-dipep-cond-dist}.
    }
    \label{fig:ala-dipep-tran-prob}
\end{figure}

In addition to the conditional distribution $P(\textbf{X}_t|\textbf{X}_0)$ of alanine dipeptide, we also estimated the transition probability $P_{\Delta t}(C_j | C_i)$, where the dihedral distribution is classified into distinct rectangular regions or classes $C \in \{\alpha, \alpha', \beta, \beta', \mathrm{other} \}$ according to Fig.~\ref{fig:ala-dipep}d (the 'other' class is the region not in $\alpha, \alpha', \beta, \beta'$). In short, $P_{\Delta t}(C_j | C_i)$ describes the probability of state $C_i$ transitioning to $C_j$ after a certain amount of time $\Delta t$. As shown in Fig.~\ref{fig:ala-dipep-tran-prob}, the transition probability matrix of SD is largely similar to that of the MD counterpart for $\Delta t = 10, 100, 200, 400$ ps, further validating the fidelity of the SD trajectory for replicating dynamic statistics.

\begin{figure}
    \centering
    \includegraphics[width=0.9\textwidth]{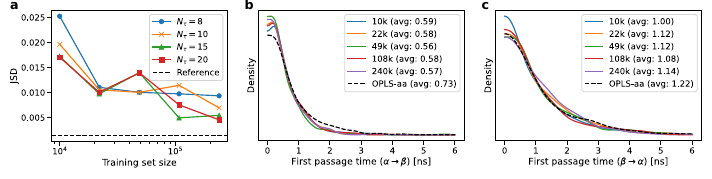}
    \caption{Convergence analysis of the alanine dipeptide equilibirum dihedral distribution and first passage time distribution with respect to training set size and number of denoising iterations.
    (a) Jensen-Shannon divergence (JSD) of the dihedral distributions between SD trajectories and the in-house MD trajectory as a function of training set size and number of denoising steps (M).
    (b, c) First passage time distribution ($\alpha \rightarrow \beta$ and $\beta \rightarrow \alpha$) measured from SD trajectories based on varying training set sizes, and from the MD reference trajectory. In (b, c), the number of denoising steps is fixed to 20.
    }
    \label{fig:ala-dipep-acc-vs-tsize}
\end{figure}

Using alanine dipeptide as a standard test molecule, a brief convergence analysis was conducted in order to investigate the SD trajectory quality with respect to two relevant quantities: the training set size and the number of denoising iterations $N_\tau$ for each SD sampling. The training set size directly dictates the amount of effort and resource one can feasibly commit for a reasonably realistic SD fitting, and the number of denoising iterations $N_\tau$ directly influences the compute cost of rolling out SD trajectories. The Jensen-Shannon divergence (JSD) of dihedral distributions between SD and MD is used as a measure of the quality of the equilibrium distribution from SD. Fig.~\ref{fig:ala-dipep-acc-vs-tsize}a shows that the JSD metric is certainly impacted by training set size, as expected, and begins to plateau after roughly $10^5$ samples. A dataset of such size is well within the capabilities of modern hardwares. On the other hand, the number of denoising iterations does not appear to influence the JSD metric in any certain fashion, i.e., more denoising iterations do not always result in lower JSD. This suggests that the inner-loop denoising ODE integration does not require a large number of iterations to be solved accurately. Therefore, for a reasonably accurate SD trajectory, one can use a fairly small number of denoising iterations, roughly in the range of 10--20, a range that is also claimed to be effective by the DPM-Solver paper \autocite{lu2022dpm}. Unless otherwise stated, we use $N_\tau=20$ throughout the paper.

Besides the quality of the equilibrium distribution produced from SD, the dynamical information, namely the first passage time distribution is also examined with respect to training set size. Fig.~\ref{fig:ala-dipep-acc-vs-tsize}bc shows that overall, the first passage time distribution is not impacted by the training set size. In (b), there is a systematic discrepancy between the SD result and the MD reference, namely the slightly faster transition rates in SD, which was previously discussed. To re-iterate here, the score model tends to underestimate the energy barriers possibly due to a lack of observations of high energy configurations in the MD-generated training dataset. We do not expect that increasing the training set size to include more high energy configurations can single-handedly and completely address the issue of energy barrier underestimation, which is indeed shown in (b). However, in (c), raising the training set size beyond 10k samples noticeably improves the first passage time distribution. Still, these improved distributions remain to have slightly faster transition rates relative to the MD reference. Future work is needed to address the issue of slightly underestimated of barriers, likely by devising more accurate neural network architectures, expanding and diversifying the training dataset to include more transitions around high energy configurations,  and/or improving the model with domain knowledge and physical constraints.

\begin{figure}
    \centering
    \includegraphics[width=0.7\textwidth]{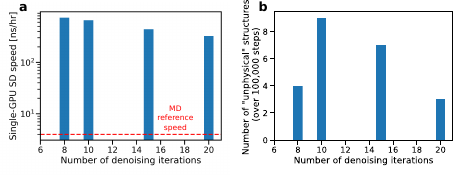}
    \caption{The impact of the number of denoising iterations on SD trajectory compute speed (a) and stability (b). The compute speed is measured in simulation time per wall time (ns/hr). The stability is measured in terms of the number of ``unphysical'' structures (those that did not satisfy the criteria detailed in Section~\ref{sec:reject-criteria}) sampled over 1 $\mu$s. In (a), the single-GPU (NVIDIA V100) MD compute speed is also provided. The SD speed is based on a NVIDIA RTX 2070S GPU.
    }
    \label{fig:ala-dipep-speed}
\end{figure}

The impact of the number of denoising iterations $N_\tau$ on SD's speed and stability for alanine dipeptide is briefly analyzed. The single-GPU SD trajectory speed is typically 80--180x faster than its MD counterpart for all values of $N_\tau$ considered in this work, and can significantly vary depending on $N_\tau$ (Fig.~\ref{fig:ala-dipep-speed}a). This result lends credence to the performance potential of SD for accelerated dynamical simulations of atomistic systems. Note that this rudimentary analysis is only based on single-GPU performance of our first Python + PyTorch implementation for a tiny molecule, which tends to involve relatively large computational overhead, compared to the highly optimized Gromacs. As part of future work, we intend to conduct a more in-depth analysis on SD performance in a multi-node, parallel setup with much larger molecules, which is commonly adopted in typical MD simulations. On the other hand, SD trajectories are generally stable. Since each timestep in SD is a probabilistic process, if an unphysical structure is encountered during SD (based on some user-defined criteria), it can simply be rejected, followed by resampling. Notably, our pre-defined criteria were mainly designed to prevent catastrophic failures such as bond breaking during SD rollouts, but such events have not been observed when the score model is well trained. The number of unphysical structures sampled over a 100,000-step rollout (a long enough time interval for statistical significance) generally is affected by the number of denoising iterations $N_\tau$, but occasionally can be low regardless of $N_\tau$ (Fig.~\ref{fig:ala-dipep-speed}b). Importantly, the rate of sampling unphysical structures is rare, and does not impact the SD compute speed. Lastly, as will be shown in Section~\ref{sec:n-butane}, the SD-sampled structures are largely physical in terms of dihedral angles, bond angles, and bond lengths. 

\subsection{Generalization to unseen n-butane} \label{sec:n-butane}

\begin{figure}
    \centering
    \includegraphics[width=1.0\textwidth]{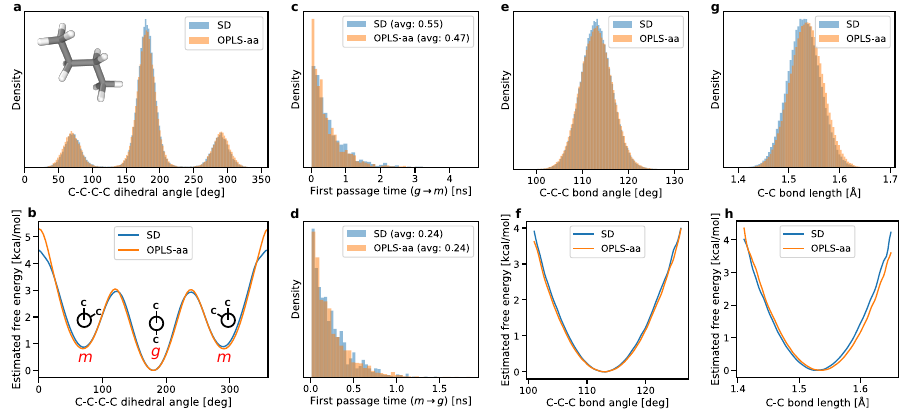}
    \caption{Score dynamics model applied to unseen n-butane, resulting in a realistic trajectory compared to in-house MD (OPLS-aa) reference. The comparison was made in terms of equilibrium distributions of the C-C-C-C dihedral angle (a, b), the first passage time distribution of the dihedral angle (c, d), distribution of all C-C-C bond angles (e, f), and distribution of all C-C bond lengths (g, h).
    The model was trained with small alkanes trajectories (ethane, propane, and pentane).
    An example configuration is visualized in (a).
    }
    \label{fig:unseen-butane}
\end{figure}

Here we demonstrate the generalizability of SD to an unseen molecule, namely n-butane. In this experiment, the score model was trained with MD trajectory data of n-ethane, n-propane, and n-pentane. After training, the model was applied to a butane configuration, which is unseen by the model. The subsequent SD trajectory over 100 ns is then analyzed and compared with a ground-truth n-butane MD trajectory of the same length (Fig.~\ref{fig:unseen-butane}), which shows that the butane SD result is highly accurate in terms of (a, b) the equilibirum dihedral distribution, (c, d) the first passage time distribution of the dihedral angle dynamics, (e, f) the C-C-C bond angle distribution, and (g, h) the C-C bond length distribution. The estimated free energy distribution plots (b, f, h) are calculated by $F = -k_B T \log{P}$, where $F$ denotes free energy, $P$ is the kernel density estimation of the corresponding marginal distribution, and $T=300$ K. Note that the high energies are slightly underestimated, similar to alanine dipeptide. Regardless, we consider the overall SD generalization result on n-butane to be accurate.

We attribute such an effective generalizability of the SD model to the node-centric nature of the underlying graph network model. Since the graph network prediction is per-atom, SD essentially amounts to predicting and sampling atomic displacements based on the local environments of each atom. In this regard, SD is expected to have similar utility and transferability of machine learning potentials, where information learned from small-scale MD data can be applied to larger, more complex systems. Future work will focus more on generalization of SD to unseen structures.  

\subsection{Generalization to out-of-distribution i-butane} \label{sec:i-butane}
\begin{figure}
    \centering
    \includegraphics[width=0.7\textwidth]{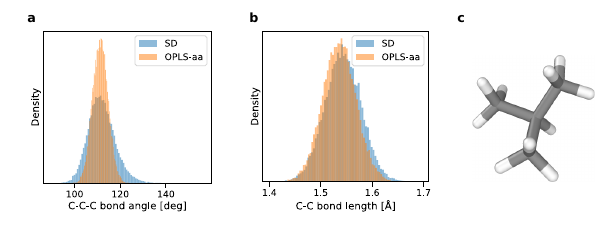}
    \caption{Score dynamics model applied to out-of-distribution i-butane, resulting in a 100 ns trajectory. (a) The equilibrium distribution of (a) C-C-C bond angles and (b) C-C bond lengths were compared to that of a reference MD (OPLS-aa) trajectory. The i-butane structure, after 100 ns of SD trajectory, is shown in (c). 
    }
    \label{fig:ood-ibutane}
\end{figure}

Here we also investigate the generalizability of SD to an unseen and out-of-distribution i-butane, in which the center carbon bonds to three methyl groups is unobserved in the training set of straight-chain n-alkanes. Regardless, the SD model was able to generate a stable trajectory over 100 ns given an initial frame of i-butane. However, the generated topology, especially the C-C-C bond angle (Fig.~\ref{fig:ood-ibutane}a), does not accurately replicate that of the reference MD trajectory. This is within reasonable expectation, as the SD model never observed the i-alkanes topology during training. On the other hand, the C-C bond length distribution (Fig.~\ref{fig:ood-ibutane}b) to a large extent is similar to the MD reference, presumably because of the abundance of C-C bonds in the training data. Overall, this result still shows promising potential for SD generalizability; we believe that with sufficient and highly varied training data, SD can generalize to a wide range of molecules. While we focus on the fundamentals and the faithful reproduction of small prototype molecules in this work, we intend to focus on generalizability and scalability in future work.

A separate test was performed for n-hexane to test the extrapolating performance. However, the results were problematic as the same SD model experienced incidents of bond breaking and could not produce long and stable trajectories. Such limited generalizability of the current SD model is motivation for future work to address this issue.

\subsection{Comparison to previous results}
We briefly compare our approach and the corresponding results to previous works. Some recent generative models were able to employ larger timesteps such as 50 ps  for the same alanine dipeptide molecule \autocite{Klein2023-Timewarp} or $\sim$ ns \autocite{Fu2022, Schreiner2023-Implicit}. Our work aims at faithfully reproducing both the equilibrium and transient properties of MD simulations. In the alanine dipeptide and small alkane examples shown above, the typical first-passage time is about 200--1000 ps, and the chosen $\Delta t=10$ ps is suitable for simulating such dihedral angle dynamics with efficiency and accuracy. In general, $\Delta t$  is determined by the time scale of the fastest degrees of freedom we care about reproducing. If one wants to accelerate the simulation with faithful thermodynamics and all kinetics except the thermal vibrations or phonon modes, then $\Delta t \sim$ps is a sensible choice.  In Ref.~\autocite{Klein2023-Timewarp}, good agreement on equilibrium distributions in Ramachandran plots were obtained, while the time correlation and jump rates were notably different from the ground truth MD, and the proposal acceptance rate was $< 2$\%. In contrast, our approach was designed to follow both thermodynamics and kinetics, with $>99.99$\% acceptance rate. Ref.~\autocite{Schreiner2023-Implicit} employed variable timesteps up to ns for alanine dipeptide and other fast-folding peptides and was therefore able to reach larger speed-up than this work, while we have focused the reproduction of low-level details such as C-C bond distance and C-C-C bond angle distribution, as well as preliminary demonstration of generalizability in alkanes.
\section{Conclusions}

This paper proposes score dynamics, a computational method to scale up molecular dynamics simulations both spatially and temporally. Rather than accelerating MD by modifying the simulated physics, SD strives to faithfully reproduce MD predictions of both kinetic processes, including transition rates and transition paths, and equilibrium distributions. This is achieved by learning the transition probability matrix between atomic configurations separated by large timesteps (ps) from MD trajectory. The transition matrix is formulated with a conditional diffusion model, whose key ingredient is a score function approximated by a graph neural network dependent on current and next configurations. In this introductory paper, we focused on detailed investigations on a few small molecules, namely alanine dipeptide and short alkanes, in aqueous solution. By taking large timesteps and omitting solvent molecules, our score dynamics approach was shown to  quantitatively reproduce both equilibrium distributions over order parameters and kinetic properties such as transition rates/first-passage time and conditional distributions of the transition path. That both equilibrium predictions derived from the stationary distributions of the conditional probability and kinetic predictions for the transition rates paths are well reproduced is a testament for the efficacy of SD. Generalization to an unseen molecule (butane) was also demonstrated. Score dynamics can be used as an accurate surrogate of high-fidelity MD method trained from MD trajectory, or applied in combination with other MD-derived methods such as replica exchange or hybrid MC/MD to achieve even larger scales. Decent wall-clock speedup (up to 180X) was obtained for the small molecules considered with inexpensive classical force field, while more sizeable gains could be achieved in larger systems using more accurate and expensive potentials with an optimized implementation. While a stretch, it might even be plausible to train SD using {\textit{ab initio}} MD without fitting interatomic potentials.

The purpose of this introductory paper is not to fully develop and refine the score dynamics method for complicated systems, but to introduce the basic framework and draw attention to related challenges in future studies. There are a fair number of open questions about SD that have been barely addressed. First, while we have demonstrated the first example of generalization in butane, extension to larger, more challenging molecules will be the subject of follow-up works. Second, the basic formalism and assumptions need a more thorough investigation. As discussed in Section \ref{sec:approx}, for example, we considered neither history dependence nor effects of velocities, simplifications that might be justifiable in the relatively simple example of small molecules in aqueous solution. For more challenging problems, memory and/or velocities should be considered. Additionally, a single timestep $\Delta t=10$ ps and a single temperature $T=300$ K were chosen. More work needs to be done to push these boundaries to reach even higher speedup or to generalize to adjustable temperature. Other issues not considered include the details of ground-truth MD simulations, e.g.\ thermostat types and parameters.

The second class of big open questions revolve around the score  or $\log P$ model. Parallels can be drawn between the importance of accurate interatomic potential models for MD, and that of score  model for SD. The former is both a well-known technical challenge and a fruitful test-bed of innovative solutions. We hope the issue of finding optimal functional approximations for the score or $\log P$ model, which directly determines the accuracy of SD, will attract interests from the community. The currently implemented score model has several obvious issues and needs to be improved. It directly predicts scores or displacement vectors for efficiency rather than derivatives of $\log P$ and is therefore not strictly conservative, similar to the difference between force-field and interatomic potential models. Future works will determine whether a conservative score model can offer better accuracy at additional computational cost. The current score model also suffers from lack of rotational equivariance and inaccuracy in its long tail of sampling rare unphysical moves (bond breaking). Similar to fitting interatomic potentials, we expect a diverse and representative training dataset, including more transitions around high energy configurations, will be beneficial for the accuracy and generalizability of fitted score models. A notable distinction from potentials is that score functions as probabilistic models require large training datasets to learn the conditional distribution. Going forward, high training cost is a major concern for future applications of SD. It is justifiable when the time savings in large or repeated SD simulations outweigh the initial training costs. Additionally, an improved model should be imbued with not only large MD data, but domain knowledge and physical constraints. Finally, quantifying the uncertainty and long-term stability of SD simulations is likely more difficult than MD. Future studies are needs to addressed these challenging problems.
\section*{Acknowledgements}
TH and FZ acknowledge support by the Critical Materials Innovation Hub, an Energy Innovation Hub funded by the U.S.\ Department of Energy, Office of Energy Efficiency and Renewable Energy, and Advanced Materials and Manufacturing Technologies Office. BS and VB are partially supported by the Laboratory Directed Research and Development (LDRD) program (22-ERD-016) at Lawrence Livermore National Laboratory. We thank Drs.\ Roland Netz and Benjamin Dalton for helpful discussions regarding butane simulations. Computing support for this work came from LLNL Institutional Computing Grand Challenge program. This work was performed under the auspices of the U.S. Department of Energy by LLNL under contract DE-AC52-07NA27344.

Supplemental information to this work is available free of charge at https://pubs.acs.org

\section*{Author Contributions}
TH implemented score model training and performed SD rollouts, with technical contribution from FZ. FZ generated the MD data, and supervised the research with inputs from all authors.

\section*{Data Availability}
The data required to reproduce this work, namely the MD dataset and the trained score model parameters, can be requested from Tim Hsu and Fei Zhou.

\section*{Code Availability}
The source code, the trained score models, and a demo for implementing the SD rollout are available at \url{http://www.github.com/llnl/graphite}.

\section*{Competing interests}
The authors state that there is no conflict of interest.

\printbibliography
\newpage
\setcounter{figure}{0}
\renewcommand{\figurename}{Supplementary Figure}
\renewcommand{\theHfigure}{Supplement.\thefigure}

\setcounter{table}{0}
\renewcommand{\tablename}{Supplementary Table}
\renewcommand{\theHtable}{Supplement.\thetable}

\appendix

\section{Ethane, propane, and pentane SD results}
The score model applied to the unseen n-butane was trained with n-ethane, n-propane, and n-pentane MD dataset, which consists of a 100 ns trajectory for each molecule. The SD performance for these alkanes---in terms of the dihedral angle, bond angle, and bond length distributions---are shown here in comparison with the corresponding MD training data. 

\begin{figure}[h]
    \centering
    \includegraphics[width=1.0\textwidth]{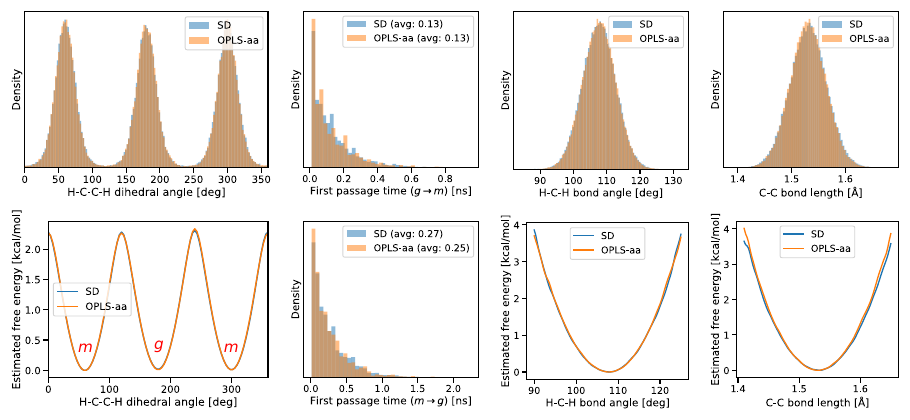}
    \caption{Ethane SD result compared to MD training data. Note that the dihedral angle distribution (and its corresponding first passage time) refers to only one of the H-C-C-H quadruplets.
    }
    \label{si-fig:ethane}
\end{figure}

\begin{figure}
    \centering
    \includegraphics[width=0.5\textwidth]{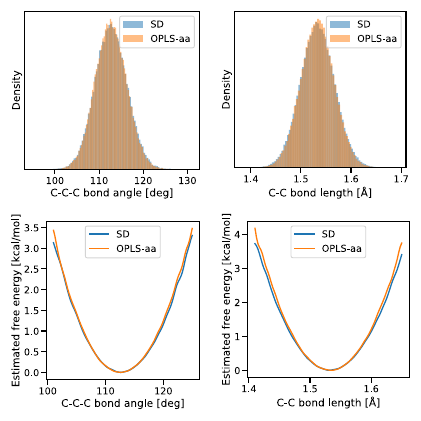}
    \caption{Propane SD result compared to MD training data.
    }
    \label{si-fig:propane}
\end{figure}

\begin{figure}
    \centering
    \includegraphics[width=1.0\textwidth]{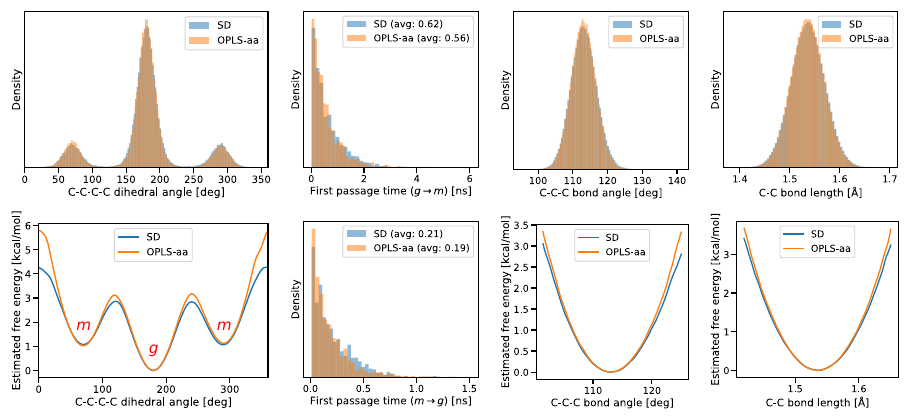}
    \caption{Pentane SD result compared to MD training data. Note that the dihedral angle distribution (and its corresponding first passage time) refers to only one of the two C-C-C-C quadruplets.
    }
    \label{si-fig:pentane}
\end{figure}

\newpage
\section{Including velocity information in the SD model}
\begin{figure}
    \centering
    \includegraphics[width=0.8\textwidth]{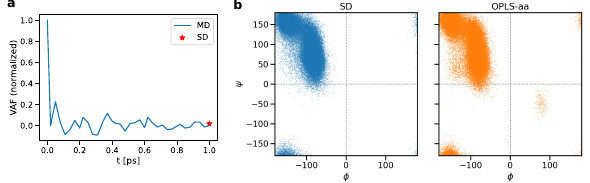}
    \caption{Velocity autocorrelation and dihedral angle distribution of alaline dipeptide MD/SD trajectories.
    }
    \label{si-fig:ala-dipep-vacuum}
\end{figure}

We performed an experiment where the velocity information is included into a separately trained position+velocity SD model, again with alanine dipeptide (this time in vacuum at $\Delta t=1$ ps interval) as the prototype test system. However, the MD trajectory of alanine dipeptide in vacuum does not have significant velocity autocorrelation after 1 ps, as shown in Fig.~\ref{si-fig:ala-dipep-vacuum}a. Therefore, the velocity information may not be very helpful to training the SD model. Regardless, preliminary analysis shows that the resulting SD trajectory accurately replicates the dihedral angle distribution of the MD reference (Fig.~\ref{si-fig:ala-dipep-vacuum}b). For a more comprehensive study, we will focus on systems and conditions with significant velocity correlation over the picosecond range in a future work.

\end{document}